# Title: Ferroelectric YMnO$_3$ films deposited on n-type Si (111) substrates

Authors: S Parashar, A R Raju P Victor and S B Krupanidhi

and C N R Rao[*]

Chemistry and Physics of Materials Unit, Jawaharlal Nehru Center for Advanced Scientific Research, Jakkur, Bangalore, India-560 064

Materials Research Center, Indian Institute of Science, Bangalore, India-560 012

YMnO$_3$ thin films have been grown on n – type Si substrates by nebulized spray pyrolysis in the Metal – Ferroelectric – Semiconductor (MFS) configuration. The C-V characteristics of the film in MFS structure exhibit hysteretic behavior consistent with the polarization charge switching direction, with the memory window decreasing with increase in temperature. The density of interface states decreases with the increase in the annealing temperature. Mapping of the silicon energy band gap with the interface states has been carried out. The leakage current measured in the accumulation region, is lower in well-crystallized thin films and obeys a space- charge limited conduction mechanism.

[*] author of correspondence; electronic mail: cnrrao@jncasr.ac.in
Fax # +91-80-8462766

Rare – earth manganates of the type $Ln_{1-x}A_xMnO_3$ where Ln is a rare earth and A is an alkaline earth have become known for a variety of fascinating properties, such as colossal magnetoresistance, charge ordering and orbital ordering [1-2]. These properties are strongly influenced by the average size of the A-site cation, $<r_A>$ [3]. The parent manganates $LnMnO_3$ also show marked effects of the size of A-site cation. $LaMnO_3$ is an antiferromagnetic insulator where the Jahn-Teller distortion around $Mn^{3+}$ plays an important role. However, as the $<r_A>$ becomes small the perovskite structure becomes unstable at ambient conditions, resulting in the hexagonal structure. Thus, $YMnO_3$ crystallizes in the $P6_3cm$ space group [4] with, a = 6.130Å and c = 11.505 Å. The structure is best described as layers of manganase-centred trigonal bipyramids with 5-coordinate Mn but no framework of Mn-O bonds along the axial direction (c-axis). The equatorial oxygens are corner-shared by three polyhedra in the basal plane. This structure leads to an unipolarization axis along [0001], which can be reversed by an electric field, thereby making these manganates ferroelectric. Although ferroelectrcity in such mangantes in the bulk form was discovered in 1963 by Bertaut et al [5], there has been little effort to prepare them in thin – film form [6] to investigate ferroelectricity. What is also noteworthy that $YMnO_3$ is also antiferromagnetic, thus making it a biferroic [7].

Interest in $YMnO_3$ and related compounds arises from their potential use in ferroelectric random access memories (FRAMs) in the metal-ferroelectric-semiconductor field-effect transistor (MFSFET) mode [8]. An additional advantage of these manganates is the absence of volatile elements such as Pb and Bi. This is especially useful in MFSFET applications as one can directly integrate onto Si substrates. In the present study, we have deposited films of YMnO$_3$ on Si (111) substrates and conducted a detailed study of their ferroelectric properties in terms of C-V characteristics.

Thin films of YMnO$_3$ were prepared by nebulized spray pyrolysis [9]. The technique involves the pyrolysis of a nebulized spray of organometallic derivatives of the relevant metals. Since the nebulized spray is deposited on a solid substrate at relatively low temperatures, with sufficient control of the rate of deposition, the oxide films obtained possess good stoichiometry. We have employed the acetate of Y and the acetylacetonate of Mn as precursors in the present study. Solutions of stiochiometric quantities of the precursors were prepared in methanol solvent. YMnO$_3$ films of ~1µm to 2µm thickness were deposited on the n-type Si(111) substrates (~$\rho$ = 1 $\Omega$ cm, dopant concentration 6.5 x $10^{14}$ cm$^{-3}$) at 450 K by using air as the carrier gas (flow rate 1.5 liters/min). The films so obtained were annealed in air at 973 K, 1073 K, and 1123 K for 6 hrs each in order to study the process of crystallization as a function of annealing temperature. The films were characterized by employing X-ray diffraction (XRD), energy dispersive x-ray analysis (EDAX) and scanning electron microscopy (SEM). For electrical measurements, sputtered gold electrodes were employed in sandwich configuration.

Fig. 1 shows the XRD patterns of the YMnO$_3$/Si (111) films annealed at different temperatures. The XRD patterns reveal that the crystallinity of the films increases as the annealing temperature is increased from 973 K to 1123 K. The evolution of the polarization axis is also seen from the increase in the intensity of the (0004) reflection. The morphology of the films was studied by SEM and the interface was found to be uniform and coherent from the cross-sectional images.

Dielectric studies and leakage current characteristics of the $YMnO_3$ films deposited on n –type Si were carried out in the Metal – Ferroelectric – Semiconductor (MFS) configuration. Gold was deposited on both the top and bottom electrodes with an area of 2 x $10^{-2}$ $cm^2$. Sweeping was carried out from the negative to a positive voltage and back, giving rise to a hysteresis behavior. Counter clockwise hysteresis was observed due to the ferroelectric polarization-switching behavior as shown in fig. 2 [10]. As the bias sweep rate was decreased, the memory window did not collapse demonstrating good memory retention characteristics of the $YMnO_3$ films on n – type Si, unlike the memory window collapse on p – type Si reported by Yoshimura et.al [6].

Polarization mode memory behavior was studied in terms of C-V hysteresis at different temperatures and results are shown in the inset of fig. 2. The polarization charge density is expected to reduce as one approach the Curie temperature, resulting in narrowing of the memory window. This behavior is clearly observed in the present studies, thereby establishes ferroelectricity in the $YMnO_3$ films.

Capacitance – voltage measurements were carried out with a computer-interfaced Agilent 4294A impedance analyzer at 100 Hz and 1 MHz with a sinusoidal voltage of 500 mv. The results are shown in fig. 3. The C – V curves at 1kHz and higher were of high frequency type with minimum capacitance at the inversion region. The minority carriers do not respond to the applied ac signal at higher frequencies in all the films and inversion layer response at 100Hz was observed in films annealed at 1123 K. This indicates that the ac signal is sufficient to induce generation and recombination of the minority carriers in the depletion layer of the bulk n–type Si, leading to charge exchange with the inversion layer (fig. 3). From the capacitance – voltage measurements, the oxide capacitance was determined by using the nonlinear least-square fit algorithm developed by Hauser and Ahmed [11]. Quantum mechanical corrections were not applied, as the films were sufficiently thick, consistent with the work of McNutt and Sah [12]. The dielectric constant obtained for the YMnO$_3$ thin films on Si in the accumulation region was 25, a value close to that reported by Yoshimura et.al [6].

The flat band voltage (V$_{FB}$) [13] was found to shift towards zero voltage with increasing annealing temperature due to better crystallinity (Table 1). The V$_{FB}$ is calculated as follows:

$$V_{FB} = \phi_{MS} - \frac{Q_I}{C_{OX}} \qquad (1)$$

with $\phi_{MS}$ denoting the work function difference between metal and semiconductor, C$_{OX}$ as the oxide capacitance and Q$_I$ the effective interface charge [14]. The interface states arise due to the dangling bonds or unsaturated bonds at the interface of Si and YMnO$_3$ thin films. The stretch out of the C–V curve at higher frequencies along the voltage axis for the YMnO$_3$/n-Si film is not appreciable, indicating the small amount of interface traps present. The interface trap density was obtained by the Castagne–Vaipalle method [15] for the films annealed at 1123 K, as they exhibited both the low and high frequency curves as shown in the fig. 3. The density of interface states was calculated from the C–V curve as follows:

$$N_{SS}(\psi_s) = \frac{1}{qA}\left(\frac{C_{OX}C_{LF}}{C_{OX} - C_{LF}} - \frac{C_{OX}C_{HF}}{C_{OX} - C_{HF}}\right) \qquad (2)$$

Here $\psi_s$ is the surface potential, C$_{ox}$ the oxide capacitance, C$_{LF}$ capacitance at low frequency (100 Hz), C$_{HF}$ capacitance at high frequency (1MHz), q charge of electron and A area of electrodes. The densities of the interface traps for films annealed at 1073 K and 973 K was calculated using high frequency Terman method [16]. The calculated values of $N_{SS}$ are shown in Table 1. The mapping of the energy band gap of Si with the density of interface states of YMnO$_3$ thin films crystallized at 1123 K is shown as an inset in fig. 3. The density of the interface states is lower than that reported earlier [6] while the $N_{ss}$ is higher at the band edges with a dip at the center of the energy band gap of Si for YMnO$_3$ films annealed at 1123 K.

Well-crystallized films of YMnO$_3$ (annealed at 1123 K) exhibited lower leakage current than those annealed at lower temperatures as shown in fig. 4. YMnO$_3$ films annealed at 1123 K obey the space-charge conduction limited mechanism (SCLC) as shown in the fig. 4. In the region I, the J–E (J is the current density in amps per unit area and E is the electric field applied) doesn't show a linear slope, instead exhibits a slope ($\alpha$) of 0.4, and in the region II ($\alpha \sim 16.1$) there is a leap in the slope at voltage trap filled limited ($V_{TFL}$) which is due to the filling up of the trap levels present in the YMnO$_3$ thin films and in the region III ($\alpha \sim 2$) it exhibits trap free square law

$$J = \left( \frac{9k\mu\varepsilon_o E^2}{8L} \right) \qquad (3)$$

Here J is current density in amps per unit area, k is the Boltzmann constant, µ is free electron mobility, $\varepsilon_o$ is the permittivity of vacuum, E is the applied electric field and L is the thickness of the film. Present observations are upto the level of indicating a trap filled space charge conduction, however a detailed conduction behavior of the YMnO$_3$ thin films on different substrates is under progress.

In conclusion, YMnO$_3$ films have been deposited successfully on n – type Si by nebulized spray pyrolysis. The bulk dielectric constant is found to be 25 from the accumulation capacitance. The C-V hysteresis behavior and the rotation direction establish ferroelctricity in these YMnO$_3$ films. The reduction in the width of memory window with ambient temperature confirms the ferroelectric behavior of the YMnO$_3$ films. The interface states have been studied and the densities of states estimated.

**Acknowledgement**

The authors thank BRNS (DAE), India, for support of this research.

**Table I**

Flat band voltages and densities of interface traps in YMnO$_3$ thin films.

| Annealing temperature (°C) | V$_{FB}$ (V) | N$_{ss}$ (eV$^{-1}$ cm$^{-1}$) |
|---|---|---|
| 700 | 0.505 | 3.2 x 10$^{12}$ |
| 800 | 0.2275 | 9.1 x 10$^{11}$ |
| 850 | 0.198 | 3.1 x 10$^{11}$ |

**References**

[1] Ramirez A P 1997 *J. Phys.: Condens. Matter* **9** 8171

[2] Rao C N R and Raveau B *Colossal Magnetoresistance, Charge Ordering and Related Properties of Manganese Oxides* World Scientific Singapore

[3] Rao C N R, Arulraj A, Cheetham A K and Raveau B 2000 *J. Phys.: Condens. Matter* **12** R83

[4] Yakel H L, Koehler W C, Bertaut E F and Forrat E F 1963 *Acta Cryst.* **16** 957

[5] Bertaut E F, Forrat E F and Fang P H 1963 *Compt. Rend.* **256** 1958

[6] Yoshimura T, Fujimura N and Ito T 1998 *App. Phys. Lett.* **73** 414 ; Kitahata H, Tadanaga K, Minami T, Fujimura N and Ito T  1999 *App. Phys. Lett.* **75** 719; Yoshimura T, Fujimura N, Ito D and Ito T 2000 *J.Appl. Phys.* **87** 3444; Fujimura N, Ito D and Ito T 2002 *Ferroelectrics* **271** 229

[7] Flebig M, Lottermoser Th., Fröhlich, Goltsev A V and Pisarev R V 2002 *Nature* **419** 818

[8] Scott J F and Araujo C A Paz de 1989 *Science* **246** 1400

[9] Raju A R and Rao C N R 1996 *Appl. Phys. Lett.* **66** 896

[10] Ito K and Tsuchiya H 1977 *Solid-State Electron.* **20** 529

[11] Hauser J R and Ahmed K edited by D.G.Sciller, A.C.Diebold, W.M.Bullis, T.J.Shaffner, R.McDonald and E.J.Walters 1998 *Characterization and Metrology for ULSI technology:The American Institute of Physics, New York* pp. 235 – 239

[12] McNutt M J and Sah C T 1975 *J. Appl. Phys.* **46** 3909

[13] Nicollian E H and Brews J R 1982 *Metal-oxide-semiconductor (MOS) Physics and*

*Technology John Wiley*

[14] Sze S M 1981 *Physics of Semiconductor devices 2$^{nd}$ ed. p 248 Wiley NewYork.*

[15] Castagne R and Vaipalle A 1997 *Surf. Sci.* **28** 557; Castagne R and Vaipalle A 1970 *C. R. Acad. Sci. (Paris)* **270**, 1347

[16] Terman L M 1962 *Solid-State Electron.* **5**, 285

**Figures**

Fig.1 X–ray diffraction pattern of the YMnO$_3$ thin films on Si(111) annealed at (a) 973 K (b) 1073 K (c) 1123 K.

Fig. 2 C-V characteristics of YMnO$_3$ film on n-type Si(111) at different bias sweep rates: 0.75 Vs$^{-1}$ (squares), 0.375 Vs$^{-1}$ (circles), 0.093 Vs$^{-1}$ (up triangles). The inset shows the C-V characteristics of the films at different temperatures. The loop area decreases as the temperature is increased: 297 K (circles), 313 K (down triangles) and 373 K (squares).

Fig. 3 C–V characteristics of YMnO$_3$ thin films on n – type Si(111) measured at 100 Hz (up triangles**)** and 1 MHz(circles**)** at 300 K. The inset shows the mapping of the density of interface states with the energy band gap of Si.

Fig. 4 Log I – Log V curves in the accumulation region in MFS structure for the different annealing temperature, .i.e. 973 K (squares)**,** 1073 K (up triangles) and 1123 K (circles).

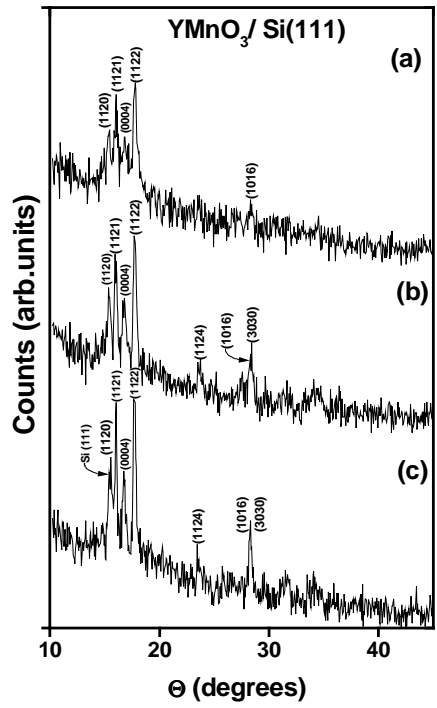

Fig. 1
Parashar *et al.*

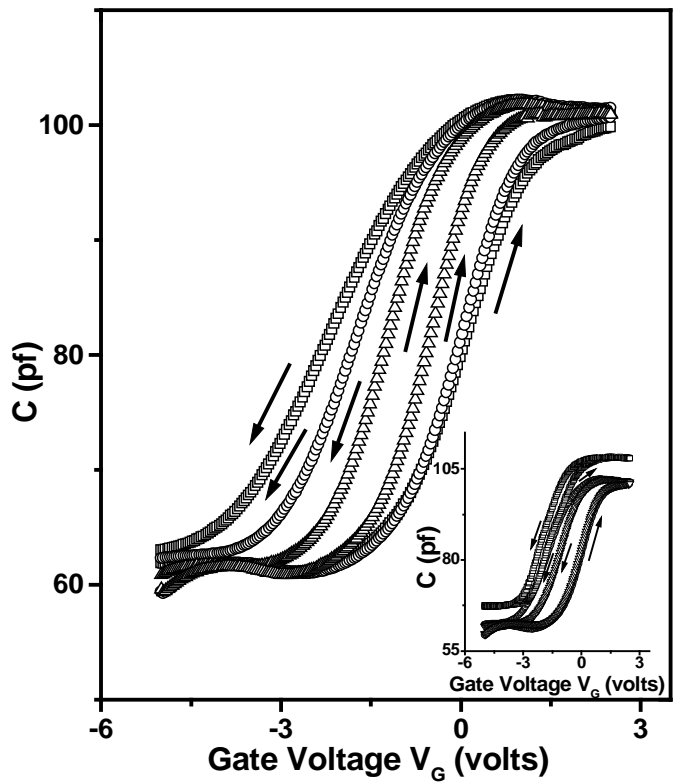

Fig. 2
Parashar *et al.*

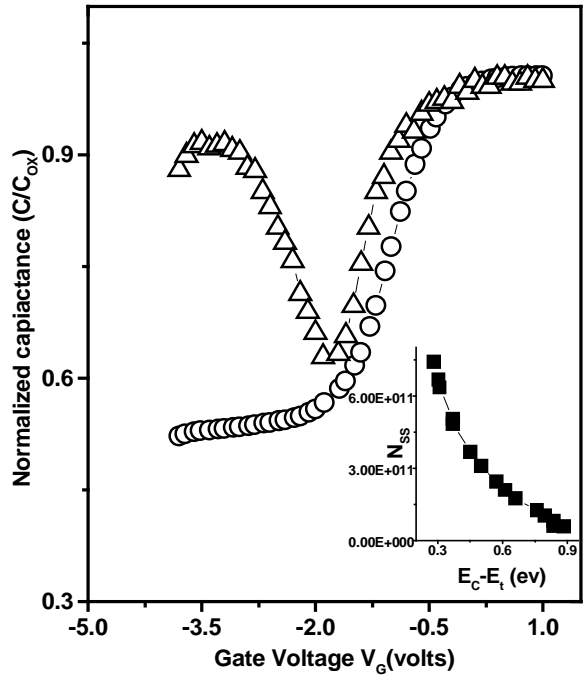

**Fig. 3**
Parashar *et al.*

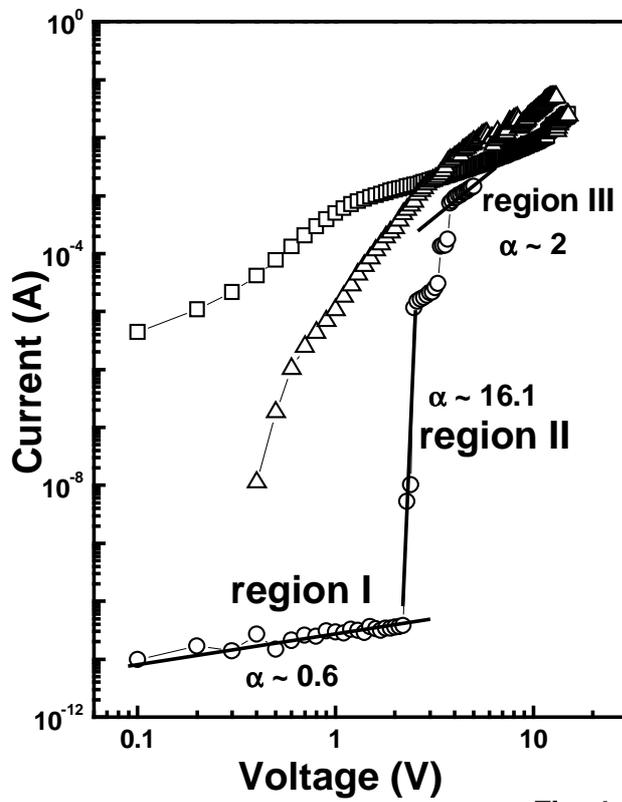

**Fig. 4**
Parashar *et al.*